# A Flat Dual-Polarized Millimeter-Wave Luneburg Lens Antenna Using Transformation Optics with Reduced Anisotropy and Impedance Mismatch

**Yuanyan Su[1], Teng Li[2], Wei Hong[2], Zhi Ning Chen[3] and Anja K. Skrivervik[1]**

[1] Microwave and Antenna Group, Ecole Polytechnique Fédérale de Lausanne (EPFL), Switzerland
[2] State Key Laboratory of Millimeter Waves, Southeast University, China
[3] Electrical and Computer Engineering Department, National University of Singapore, Singapore

Abstract— In this paper, a compact wideband dual-polarized Luneburg lens antenna (LLA) with reduced anisotropy and improved impedance matching is proposed in Ka band with a wide 2D beamscanning capability. Based on transformation optics, the spherical Luneburg lens is compressed into a cylindrical one, while the merits of high gain, broad band, wide scanning, and free polarization are preserved. A trigonometric function is employed to the material property of the flattened Luneburg lens with reduced anisotropy, thus effectively alleviates the strong reflection, the high sidelobes and back radiation with a free cost on the antenna weight and volume. Furthermore, a light thin wideband 7×1 metasurface phased array is studied as the primary feed for the LLA. The proposed metantenna, shorted for metamaterial-based antenna, has a high potential for B5G, future wireless communication and radar sensing as an onboard system.

## 1. INTRODUCTION

Multibeam and beamscanning antennas are increasingly demanded in wireless communications and remote sensing in millimeter-wave (mmW) bands, as they enhance the signal-to-noise ratio by agilely radiating toward the receivers with a highly directive beam and increase the diversity through the multipath effects. A conventional way to realize beam scanning is to utilize phased array antennas with dedicated beamforming networks (BFNs). However, the corresponding RF electronics to control the magnitude and phase are costly and high-loss in mmW bands, and the footprint of these RF electronics would hamper the system miniaturization. Alternatively, based on ray optics, lens antenna is a good candidate to realize the beamscanning functionality by use of a few spacial feeds without blockage problems. Even without BFNs, it still can easily achieve a good scanning coverage by mechanically steering its feeding antenna or switching its spacial feeds. Moreover, the lens itself is essentially frequency-independent, which hence features a broadband operation.

Among the various types of lenses, Luneburg lens antenna (LLA) has attracted more attention in the recent years because it can achieve a full ± 180° beam coverage with zero scanning loss, high gain, high efficiency and polarization independence. However, its unique sphere geometry and gradient indexes increase the complexity of antenna fabrication, lens installation, and system deployment particularly in the low frequency bands. To tackle this, transformation optics (TO) which is able to compress LL into an arbitrary shape and preserve the original LL's functionalities [1, 2], is rapidly developed with the gradient-index metamaterials by manipulating the transformed material properties [3-7]. This method has brought a new way to realize beamscanning/multibeam antennas for future communications. Several attempts have been found in the open literature to improve LL so as to address some critical issues for mmW applications. For instance, an anti-reflective layer attached on the flat surface of the quasi-conformal transformed LL is proposed to solve the reflection and beam broadening problem [8]. A waveguide half-Luneburg geodesic lens with the modulated geodesic profile in [9] enables the in-plane size reduction. An ellipsoid lens antenna compressed from LL using TO has been presented in [10] to achieve a wide-scan ability, but the bandwidth is limited due to its microstrip feeding array. A wideband printed-circuit-board (PCB)-stacked Luneburg reflector lens proposed in [11] has a low-profile strength, but it also suffers from the narrow 1D scanning range. A compact 3D-printed LL with a high compression ratio shown in [12] has been realized using five different dielectric filaments, which is then mechanically fed by an open-ended waveguide. However, obtaining an excellent overall performance in terms of radiation, bandwidth, polarization, volume, beam coverage, fabrication, and system deployment is still a challenge at present.

In this paper, a new compact PCB-stacked dual-polarized LL based on TO is proposed to achieve a broad 2D beam coverage and a small $F/D$ ratio of 0.43 in Ka band for the convenience of mass fabrication. To mitigate the impedance mismatch and wave reflection on the lens surface, the transformed material properties after anisotropy reduction can be modified with two additional design degrees of freedom by introducing a trigonometric function. Meanwhile, to acquire the electrical beamscanning capability, a low-profile wideband 7×1 metasurface antenna array is optimized as the primary feed to enlarge the beam coverage, avoid the grating lobes, and reduce the mutual coupling and pattern distortion for the high-gain onboard LLA system.

The paper is organized as follows. The design principle of the flattened LL is briefed in Section 2 with the fundamentals of TO method. The impedance mismatch reduction method to the transformed material profile of the flat LL is also described in this section, associated with a discussion on the performance improvement. Section 3 demonstrates the implementation of the flattened LL design, the metasurface feeding antenna array, and the entire onboard LLA in mmW band, the performance of the latter is presented at the end of this section. The fabrication prototypes and the measurements are presented in Section 4 to validate the design, which is followed by the conclusion in Section 5.

## 2. DESIGN PRINCIPLE

### (1) Flat Luneburg lens with transformation optics and reduced anisotropy

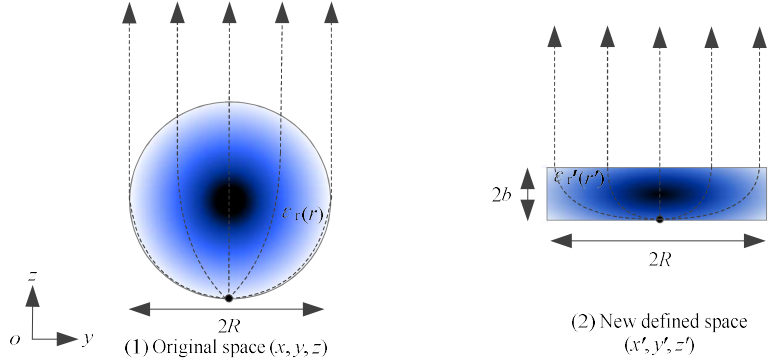

Fig. 1 Demonstration of transformation-optics principle to compress a spherical LL to a cylinder LL.

As illustrated in Fig. 1, the purpose of using TO method is to reshape the LL from an old space $(x, y, z)$ to a new defined space $(x', y', z')$, e.g., from a sphere with a radius $R$ to a thin cylinder with a thickness of $2b$ in our case, and to keep the electromagnetic wave propagation unchanged outside the lens structure. The geometry transformation mapping [6] can be mathematically expressed as

$$\begin{cases} z' = \frac{b}{\sqrt{R^2 - y^2}} z \\ y' = y \end{cases} \tag{1}$$

It reveals that the sphere is longitudinally compressed along $z$ axis, while maintaining the same scale along the horizontal $y$ axis in the new space. Since Maxwell's equations are form-invariant and same boundary conditions are remained after this transformation, the same functions of the spherical LL can be accomplished in the flattened LL by simultaneously converting the original gradient index profile of $n(r) = \sqrt{2 - (\frac{r}{R})^2}$ into an anisotropic inhomogeneous material characterized by the following equations:

$$\bar{\bar{\varepsilon}}' = \frac{\bar{\bar{\Lambda}} \, \bar{\bar{\varepsilon}} \, \bar{\bar{\Lambda}}^T}{det(\bar{\bar{\Lambda}})} = \varepsilon \left( \sqrt{y'^2 + z'^2} \right) \times \frac{\sqrt{R^2 - y'^2}}{b} \begin{bmatrix} 1 & 0 & 0 \\ 0 & \frac{-B + \sqrt{B^2 - 4C}}{2} & 0 \\ 0 & 0 & \frac{-B - \sqrt{B^2 - 4C}}{2} \end{bmatrix} \tag{2-1}$$

$$\bar{\bar{\mu}}' = \frac{\bar{\bar{\Lambda}} \, \bar{\bar{\mu}} \, \bar{\bar{\Lambda}}^T}{det(\bar{\bar{\Lambda}})} = \frac{\sqrt{R^2 - y'^2}}{b} \begin{bmatrix} 1 & 0 & 0 \\ 0 & \frac{-B + \sqrt{B^2 - 4C}}{2} & 0 \\ 0 & 0 & \frac{-B - \sqrt{B^2 - 4C}}{2} \end{bmatrix} \tag{2-2}$$

where $\bar{\bar{\varepsilon}}$ and $\bar{\bar{\mu}}$ are the permittivity and permeability in the old space, $B = -\left[ b^2 (R^2 - y'^2)^{-1} + 1 + z'^2 y'^2 (R^2 - y'^2)^{-2} \right]$, $C = b^2 (R^2 - y'^2)^{-1}$, and $\bar{\bar{\Lambda}} = \partial(x', y', z') / \partial(x, y, z)$ is Jacobian matrix that relates the coordinate transformation with the material parameter transformation.

The transformed material profiles of the flattened LL when $R = 32$ mm and $b = 4$ mm are plotted in Fig. 2 for transverse electric (TE) and transverse magnetic (TM) polarization, respectively. As can be seen, the anisotropic material contains three diagonal components along $x$-, $y$-, $z$- directions. The components of $\varepsilon_{xx}$, $\mu_{yy}$, and $\mu_{zz}$ are in response to the TE polarization (or $x$-polarization), while the components of $\varepsilon_{yy}$, $\varepsilon_{zz}$, and $\mu_{xx}$ have response to the TM polarization (or $y$-polarization). The similarity between $\varepsilon_{xx}$ and $\varepsilon_{yy}$ enables an easy implementation for dual polarization. However, it is clearly observed that due to the gradient $\mu_{yy}$ and $\mu_{xx}$, this kind of material is too sophisticated to be realized even using the metamaterials. Therefore, the anisotropy reduction which is equalizing $\varepsilon_{xx}$ and $\varepsilon_{yy}$ and unifying all permeability components, is necessary so as to simplify the transformed material for the later implementation. The selection rule between $\varepsilon_{xx}$ and $\varepsilon_{yy}$ for the final quasi-isotropic material is based on the dual-polarized radiation performance of the constructed lens. According to our study, $\varepsilon_{yy}$ performs similar to $\varepsilon_{xx}$ in these dimensions in terms of the antenna gain, sidelobe level (SLL) and scanning range for both TE and TM polarizations, it is thus selected for the next process.

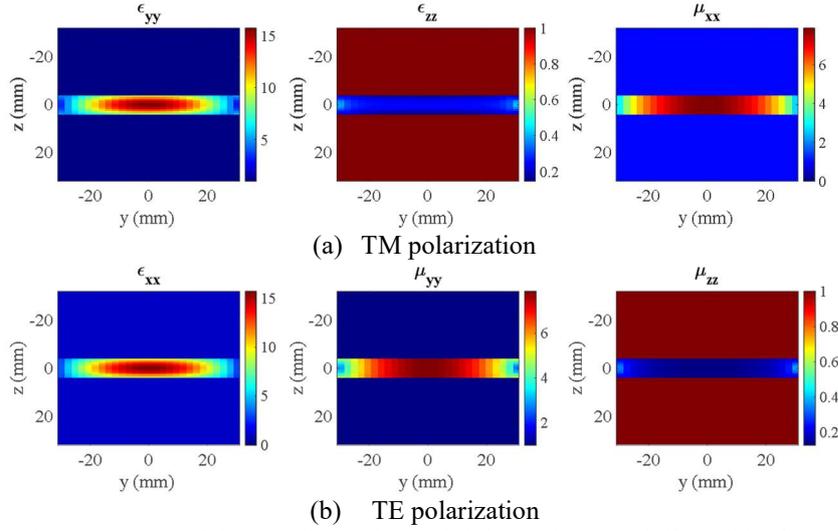

(a) TM polarization

(b) TE polarization

Fig. 2 Three diagonal components of the transformed permittivity and permeability of the flattened LL for TM and TE polarization, respectively.

(2) *Impedance mismatch reduction*

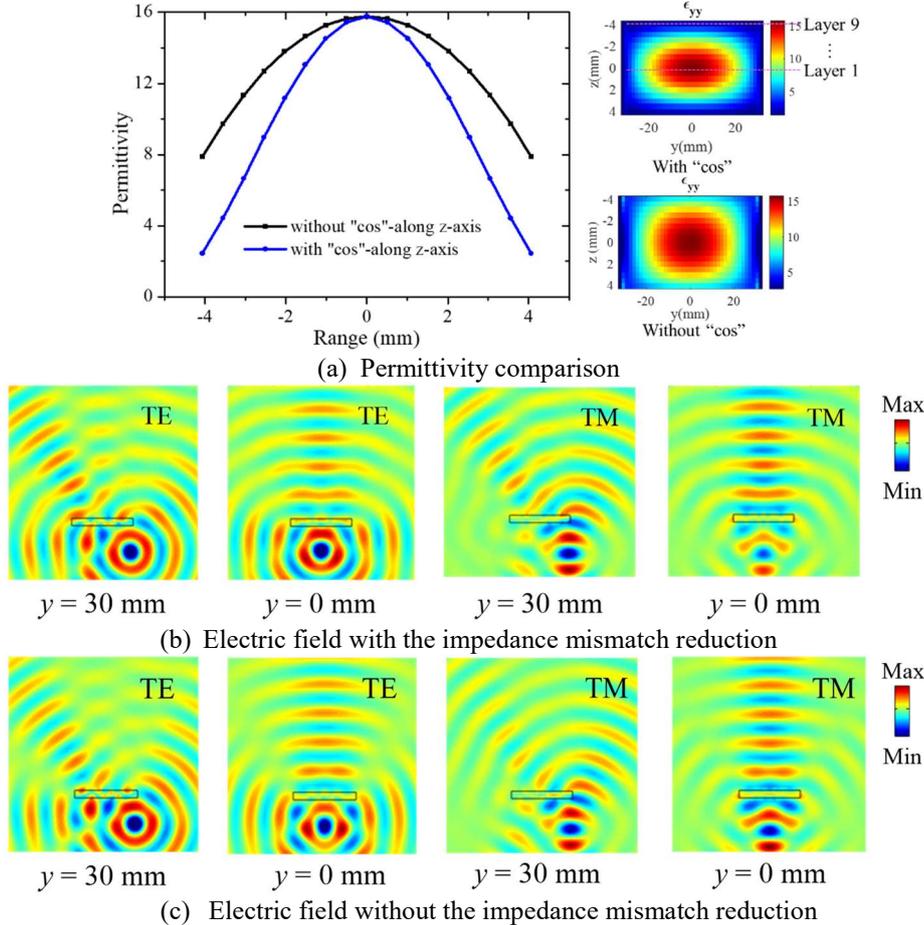

(a) Permittivity comparison

(b) Electric field with the impedance mismatch reduction

(c) Electric field without the impedance mismatch reduction

Fig. 3 Comparison of permittivity and electric field distribution with and without the impedance mismatch reduction in TE and TM polarization cases.

The feasible PCB-based implementation method proposed in [6] could benefit from the aforementioned anisotropy reduction. However, it is found that the difference between the permittivity and permeability at the lens's surface is dramatically increased, thus results in the impedance mismatch to the free-space one and the high reflection on its surface. In particular, when the compression ratio is high for a thin LL, this impedance mismatch will be severer and deteriorate the lens performance significantly. To overcome this drawback, several mitigation techniques in the literature might be investigated and applied to the flattened LL. For example, quarter-wave matching layers or transformers such as cascaded matching layers, corrugated surfaces, and pattern arrays can be attached on the lens surface [13-17]. Reactive wall might be inserted into the lens structure [13]. Also, the matching layer [18-20] and fixed anti-reflection layer [21-22] could be considered as a co-design and integrated into the gradient-index lens. One common weakness of these techniques is that the extra matching structures would increase the dimensions and the weight of the LLA, and the constraints from the matching structures may lead to potential deterioration on the lens performance.

In this regard, a direct method where a trigonometric function is applied to weight the transformed permittivity of the flattened LL, is investigated such that the permittivity gradually decreases and approaches the free-space one at the lens surface, while the permeability is simply isotropic and homogeneous. The specific trigonometric function is expressed as $a \times \cos(\frac{\pi}{d}z')$, where $a$ and $d$ are the parameters to be optimized, and the resultant weighted permittivity profile of the flattened LL now becomes

$$\varepsilon_{yy} = a \times \cos\left(\frac{\pi}{d}z'\right) \times \varepsilon\left(\sqrt{y'^2 + z'^2}\right) \times \frac{\sqrt{R^2 - y'^2}}{b} \times \frac{-B + \sqrt{B^2 - 4C}}{2} \quad (3)$$

In this way, two additional design degrees of freedom are introduced. The theoretical analysis and the optimization are conducted with the Finite Element Method (FEM) in *COMSOL Multiphysics*. The optimized parameters of $a$ and $d$ are 1 and 10, and the corresponding permittivity distribution is plotted in Fig. 3(a). Compared to the transformed permittivity without this *cos*-weighting function, the new permittivity gradually decreases from 15.7 at the lens center to 1 at the lens surface, while the maximum permittivity remains identical to the one without weighting. Therefore, it does not increase any fabrication difficulty to achieve a high permittivity. In addition, the strong reflection occurred at the lens surface is further alleviated so that it less distorts the feeding field pattern, as evidenced in the electric field distribution in Figs. 3(b-c). The planar wavefront as well as the field strength that is converted from the incident spherical wave from a feeding line source, can be slightly improved as well. When it emits out of the flattened LL, a more directive beam or a higher gain can be achieved in the farfield radiation at that angle can be achieved. It is deserved to note that the proposed method can be employed to a different structure using an arbitrary transformation mapping, which has been verified in theory using a different TO mapping but skipped here for brevity. However, this weighting function can only be applied to the transformed permittivity with the anisotropy reduction, it is not directly applicable to the anisotropic transformed material as expressed in Eq. (2).

(3) *Radiation improvement*

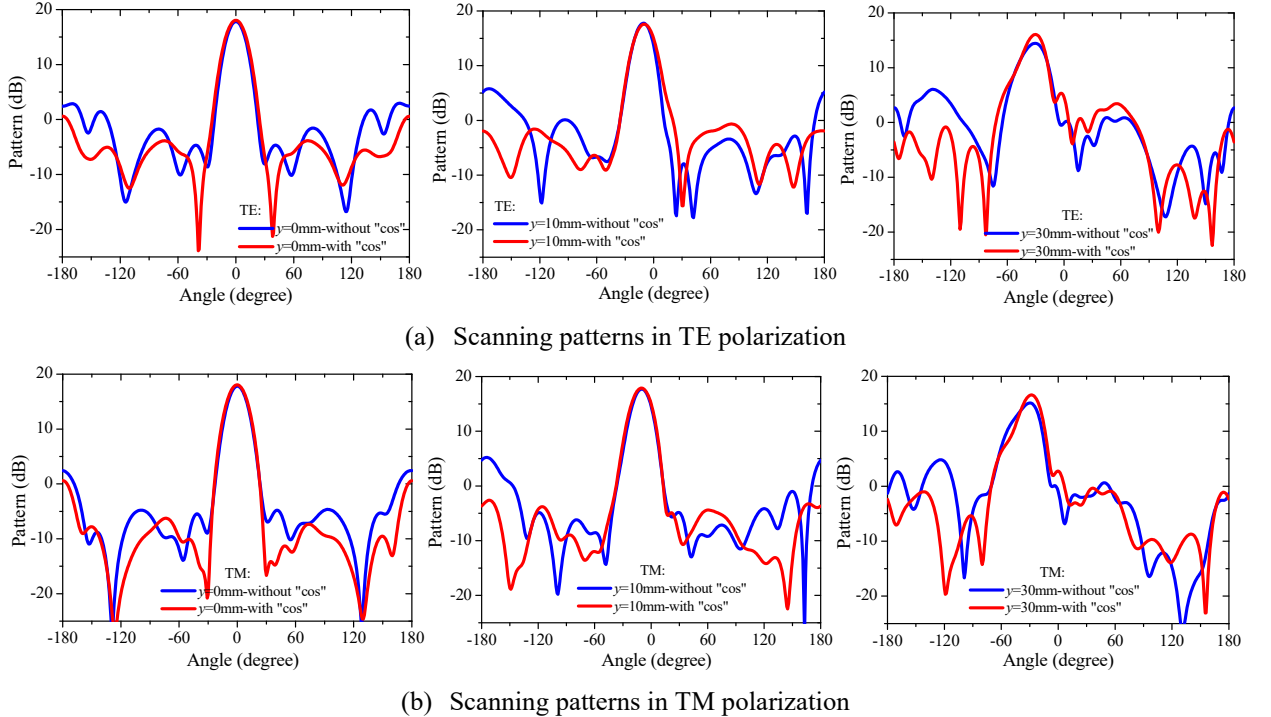

(a) Scanning patterns in TE polarization

(b) Scanning patterns in TM polarization

Fig. 4 Scanning pattern comparison between the flattened LLs with and without the impedance mismatch reduction in TE and TM polarizations, when the feeding antenna moves off the boresight to $y$=0, 10, 30 mm.

In order to demonstrate the farfield radiation performance improvement of the flattened LL with the impedance mismatch reduction, a thin LL operating at 10 GHz in X band is evaluated for dual polarization here and a stacked aperture coupled patch antenna is used as its feeding antenna, compared to the reference design in [6]. The lens dimensions used here are same as the reference one for a fair comparison. The feeding antenna is mechanically steered along a straight trajectory line specified by the focal plane. Fig. 4 plots the scanning patterns of the flattened LLA before and after applying the trigonometric weighting function. As expected, no matter for the TE or TM polarized incident wave, the sidelobes and back radiation due to the strong reflection are effectively suppressed when the beam is steering from 0° to 30°. Also, the antenna gain in the oblique incidences is further increased, while the scanning angle can be maintained. Overall, these improvements have evidenced the effectiveness of the proposed trigonometric weighting function, with a free extra cost on the weight, size, fabrication, and assembling of the entire system.

## 3. MILLIMETER-WAVE IMPLEMENTATION DESIGN AND PERFORMANCE

### (1) *Implementation of flattened Luneburg lens*

The flattened LL proposed in Section 2 is implemented in this section. For the sake of a simple manufacture with PCB techniques, the continuous permittivity profile in theory should be discretized into several thin layers with the standard PCB thickness of 0.508 mm. Therefore, 17 layers are stacked in total along $z$ direction. The permittivity of each 2D layer in *x-o-y* plane formed by 41×41 pixels is then determined based on the distribution of Layer $i$ ($i = 1, \ldots, 9$) in the *y-o-z* cross section, as shown in Fig. 5 for the half of the flattened LL that is symmetric with respect to $y$ axis.

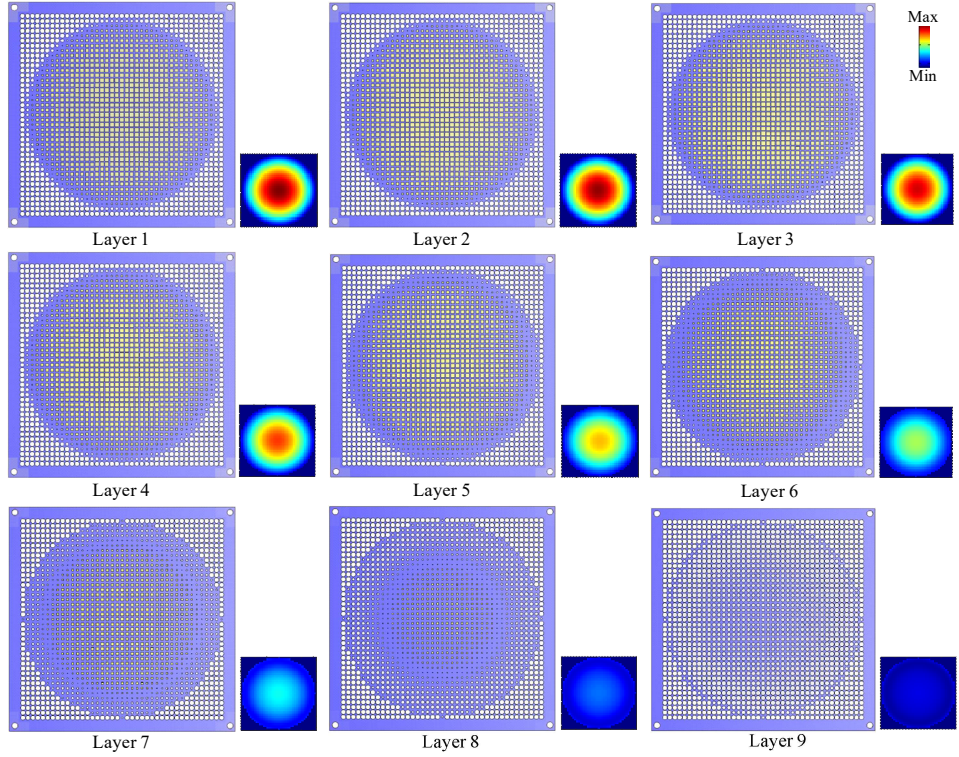

Fig. 5 Permittivity distribution and unit-cell implementation of 2D discretized layers in *x-o-y* plane for the half of the flattened LL.

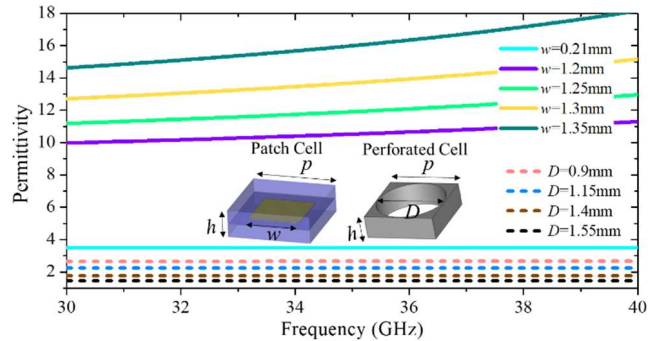

Fig. 6 Unit cell designs and their extracted non-resonant effective permittivity versus frequency.

Since each pixel represents a constant permittivity in a macroscopic view, a large dynamic range for the gradient permittivity is desirable. To achieve this goal, effective permittivity and effective permeability extracted from two kinds of unit cells with a periodicity of $p = 1.6$ mm are utilized based on the effective medium theory [23]. As presented in Fig. 6, the square patch sandwiched with two identical dielectrics is used to realize the high permittivity ranging from 3.5 to 16, while the square dielectric perforated with a circular hole at its center is used to obtain the low permittivity range of $1.45 \sim 3.4$. The dielectric substrate used here is Rogers 4003C with a dielectric constant of 3.38 and a tangential loss of 0.0027 at 10 GHz. By varying the patch size and the hole size, the stable effective permittivity and effective permeability over a broad non-resonant region of 30~40 GHz (at least) can be achieved, thus guaranteeing a wideband performance for the constructed lens. These periodic unit cells are then arranged and stacked together as 17 gradient-index metamaterial layers and finally constructs the whole flat LL.

(2)  *Metasurface feeding antenna array*

(a)  Configuration

To realize a compact antenna system and reduce the mechanical complexity, a 7×1 low-profile, light-weight, wideband metasurface antenna array is proposed as the primary feeding antenna for the flattened LL, associated with a high polarization isolation between TE and TM polarizations. The configuration of the proposed design is illustrated in Fig. 7. Each single element comprises a 2×2 metasurface with an overall size of 3.37×3.37 mm². Four degenerate modes from the metasurface can be well excited by the use of a differential microstrip-fed patch beneath it, which is then backed with a large ground plane. The distance between the upper metasurface and the driven patch is 0.508mm, a thin thickness of the loaded Rogers 4003C substrate. The truncated four corners of the driven patch are used to improve the impedance matching of the metasurface antenna.

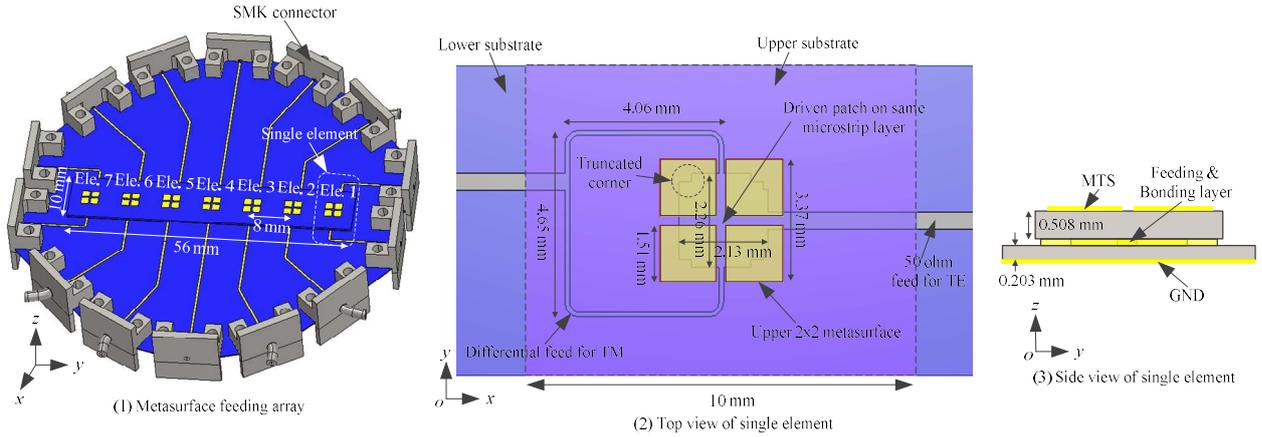

Fig. 7 Configuration of the proposed 7×1 metasurface feeding antenna array.

Benefited from the differential feeds, the isolation between the two orthogonal polarizations can reach the highest 40 dB within the bandwidth of interest, as shown in the simulated S-parameters in Fig. 8(a). An overlapped 28.6% input impedance bandwidth starting from 30 GHz to 40 GHz is obtained, and a broader bandwidth can be achieved for TM polarization. The farfield radiation of the single element at the center frequency of 35 GHz is depicted in Fig. 8(b), showing nearly-symmetric patterns with broad beamwidth and low backlobes in the scanning plane ($\varphi = 0°$) for TE and TM polarizations. It is noticed that the ideally symmetric patterns in the plane of $\varphi = 90°$ [24] are now slightly distorted and this distortion is mainly caused by the asymmetric feedings and the larger ground size in that plane for the connectors. Nevertheless, a broad beamwidth in the scanning plane is successfully achieved for the linear feeding array such that the aperture of the whole lens can be fully utilized. The element spacing is then optimized after the feeding array is integrated with the flattened LL.

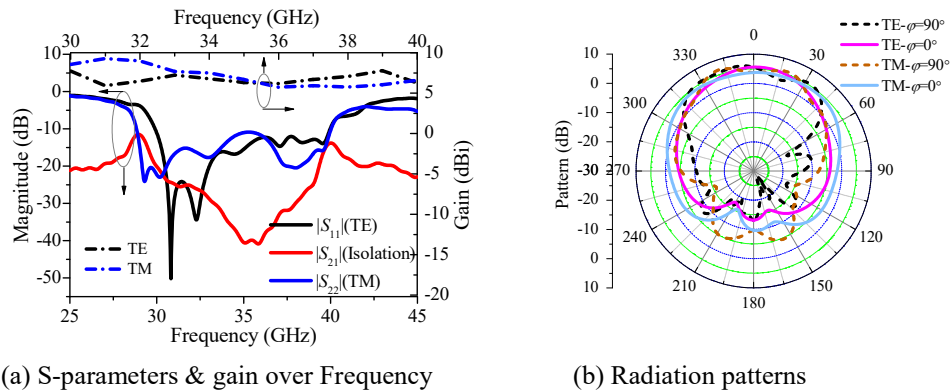

(a) S-parameters & gain over Frequency    (b) Radiation patterns

Fig. 8 Simulated performance of the proposed metasurface antenna element with SMK connectors.

(b)  Characteristic mode analysis to the wideband mechanism

The wideband mechanism of the proposed single-layer metasurface with 2×2 square patches can be interpreted by using the characteristic mode analysis (CMA) method. Fig. 9 exhibits the modal significance of this metasurface plotted in *CST studio suite*, indicating the number of modes of interest within the operation bandwidth. As can be seen, there totally exists 8 different modes from 30 GHz to 40 GHz, i.e., $J_1 \sim J_8$. In particular, Modes $J_1/J_2$ and Modes $J_7/J_8$ are two pair of (quasi-) degenerate modes that interest us most, because they are able to produce the desirable broadside radiation. As shown with the character currents and radiation pattern for each mode in Fig. 9, the character currents with respect to the mode are in-phase on the metasurface, thus contribute to the maximum gain at the boresight direction. Besides, Mode $J_1$ and Mode $J_2$ has a 90° angular phase difference and so does another pair of Modes $J_7/J_8$, which enables an excellent dual-polarized performance. As a result, combining these modes could obtain a wide operation bandwidth. Last but not least, since the rest of the modes contain anti-phase character currents and generate a null at the boresight of the character radiation pattern, they are not considered in the design.

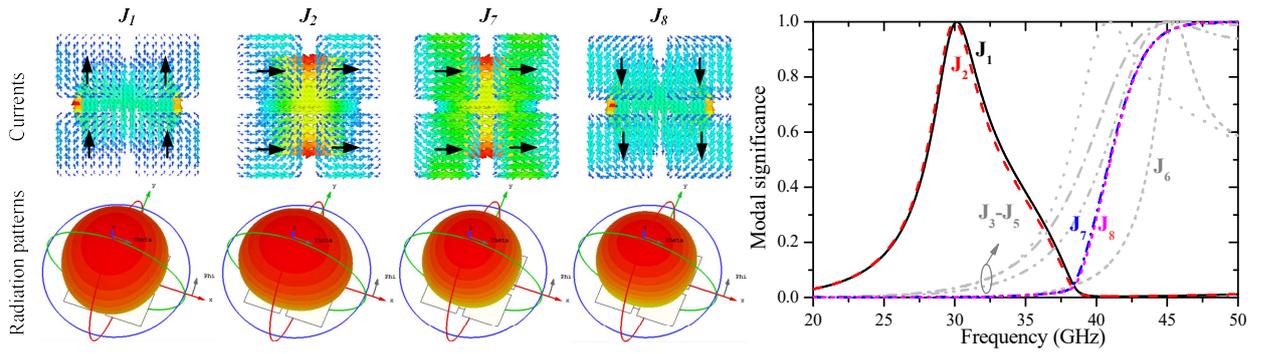

Fig. 9 Modal significance of the proposed single-layer metasurface antenna and the corresponding character currents and radiation pattern for each mode within the bandwidth of interest.

(3) *Integrated LLA and performance discussion*

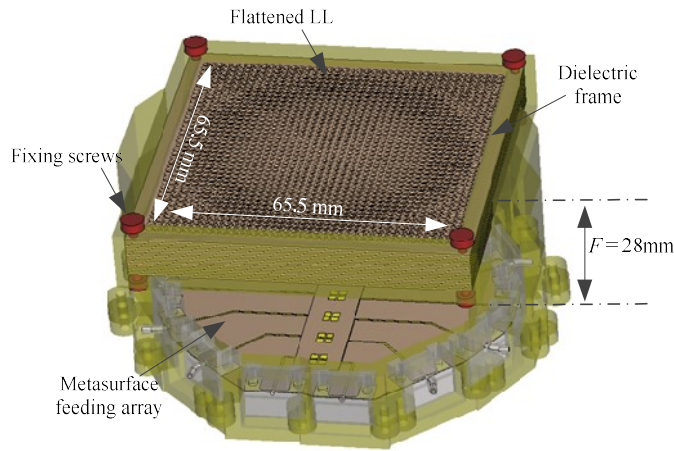

Fig. 10 Configuration of the proposed flat LLA, integrating the flattened LL and the metasurface feeding array.

Combining the proposed flat LL and thin metasurface feeding array, a very compact LLA with the optimized focal length of 28 mm is eventually obtained. The configuration of the whole LLA is detailed in Fig. 10. The $0.88\lambda_{0(33\text{GHz})}$ element spacing ($= 8$ mm) is optimized to enlarge the beam coverage, avoid grating lobes, and reduce mutual coupling and pattern distortion for the LLA. In this way, seven beams in total are used to demonstrate the scanning ability of the proposed LLA. The $F/D$ ratio of the proposed LLA is merely 0.43 and its overall aperture size is $7.2\lambda_0 \times 7.2\lambda_0$.

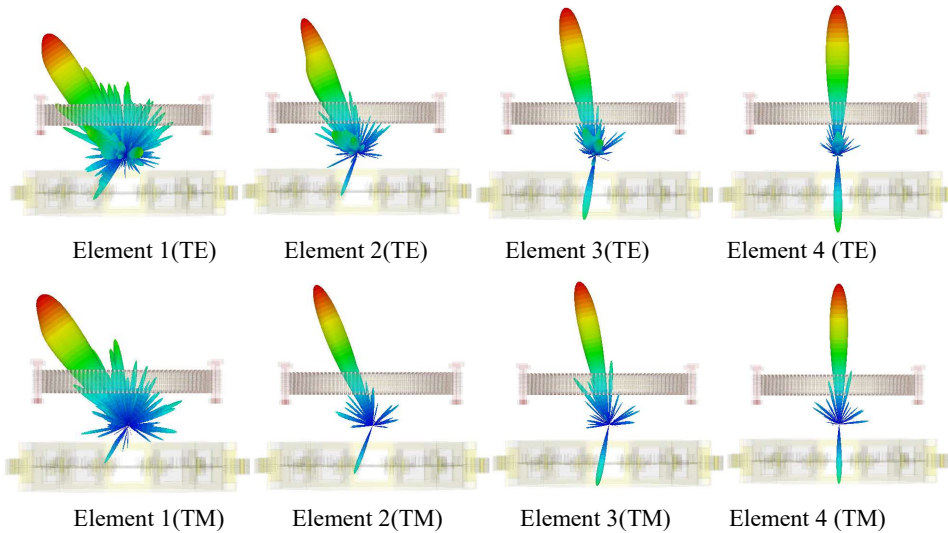

Fig. 11 3D scanning patterns of the implemented flat mmW LLA in TE and TM polarizations at 32 GHz.

The scanning performance of the implemented flat mmW LLA is examined in Fig. 11 with its 3D radiation patterns at an example frequency of 32 GHz in TE and TM polarizations. As can be seen, when the elements of the metasurface feeding antenna array are sequentially activated from the right to the left side, the antenna mainbeam is tilted to the angle of $-34°$, $-22°$, $-12°$, $0°$ with respect to the boresight direction for both two polarizations. This is in good agreement with the theoretical results using the continuous permittivity profile in Section 2(2). Therefore, it is verified that the discretization in the proposed implementation method has little negative impact on the scanning performance of the flattened LL. The maximum antenna gain occurring at the 0° scanning angle is 21.8/22.4 dBi with an aperture efficiency of 36.3%/43.7% and the half-power beamwidth reaches 9°/ 7.7° for TE/TM polarization correspondingly. The scanning loss is less than 4 dB in general and the cross-polarization level over the scanning range is < −15.5 dB and < −20.9 dB for TE and TM polarization, respectively.

## 4. FABRICATION PROTOTYPES AND MEASUREMENT VALIDATION

### (1) *Fabrication Prototypes*

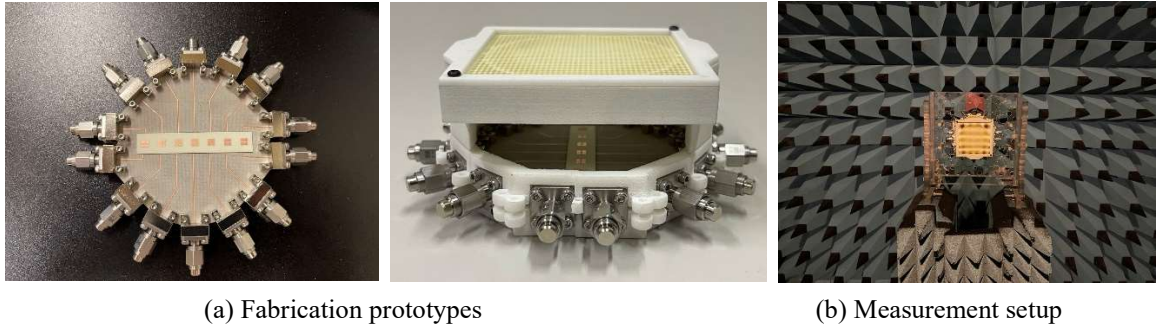

(a) Fabrication prototypes    (b) Measurement setup

Fig. 12 Fabrication prototypes of the developed compact mmW LLA and its metasurface feeding array and the measurement setup for the farfield patterns in an anechoic chamber.

The fabrication prototypes of the flat PCB-stacked LL, metasurface feeding antenna array, and the integrated LLA are presented in Fig. 12(a). All layers of the flattened LL and the feeding array can be simply fabricated using the standard PCB techniques, thus bringing a high potential for the mass fabrication. The flattened LL and the feeding antenna array are easily assembled with a dielectric frame 3D-printed by SLS (Selective Laser Sintering) technique. The S-parameters are measured using a PNA network analyzer E8361A operating up to 67 GHz. The farfield scanning patterns are measured in an anechoic chamber where the measurement setup is shown in Fig. 12(b).

### (2) *Measurements and performance*

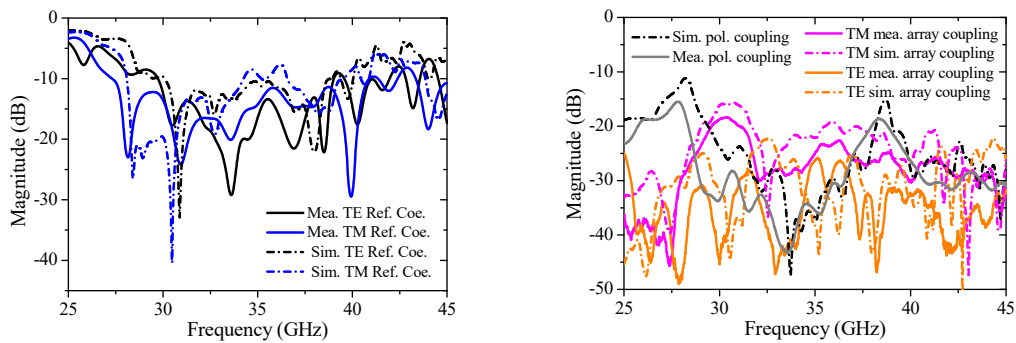

(a) Reflection coefficients for dual polarization    (b) Polarization coupling and array coupling vs. frequency

Fig. 13 Comparison between simulations and measurements in terms of the reflection coefficients, polarization coupling, and array coupling versus frequency in TE and TM polarizations.

Fig. 13(a) compares the simulated and measured reflection coefficient bandwidths of the integrated LLA when the center element is excited. It is shown that the measurement results are comparable to the simulation ones and the LLA has the wider bandwidth in TM polarization than in TE polarization, which shows its similarity with the single metasurface feeding element alone in Fig. 8(a). The overlapped impedance bandwidths of the entire LLA for the tested seven elements, which are 30.8~38.9 GHz and 27.6~38.8 GHz for TE and TM polarizations, also reasonably match with the corresponding feeding array ones. Compared to the single feeding element alone, the slight bandwidth reduction of LLA is mainly caused by the close distance from the edge array elements to the SMK connectors. Fig. 13(b) plots the worse isolation performance between two adjacent elements for the two orthogonal polarizations. It is revealed that the strongest mutual coupling happens between Element 4 and Element 3, while this coupling of < −25.8 dB in TE polarization is much lower than that of < −18.3 dB in TM polarization. The polarization coupling for each array element can be less than −30 dB on average, which enables a good dual-polarized performance.

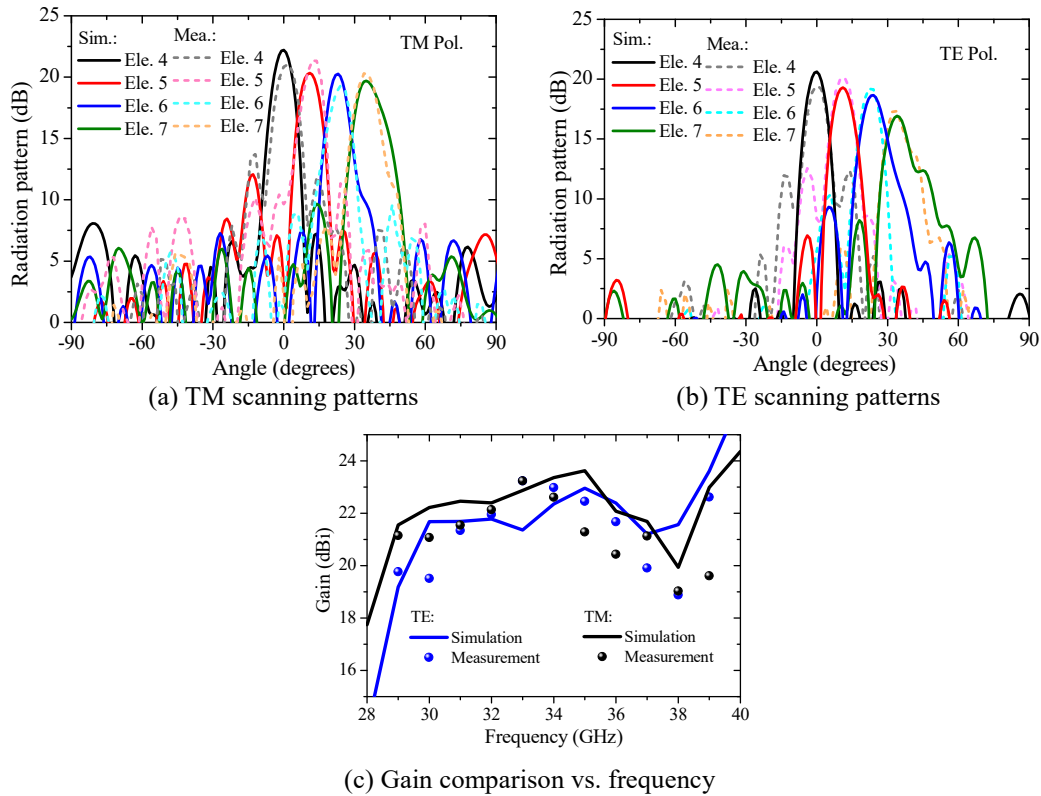

(a) TM scanning patterns

(b) TE scanning patterns

(c) Gain comparison vs. frequency

Fig. 14   Measured scanning patterns and boresight gain of the proposed LLA in TE and TM polarizations at 30 GHz, compared to the simulation ones.

From the gain measurement in Fig. 14, we can see that the 3dB gain bandwidth is slightly shifted down and narrower than the impedance bandwidth. The measured and simulated scanning radiation patterns of the proposed LLA at 30 GHz are exhibited in Figs. 14(a-b) for TM and TE polarizations, respectively. Due to the symmetry of the LLA, only the right half of the scanning plane is plotted for brevity. As can be seen, the proposed LLA is able to achieve a scanning range of ±34° and ±35° for TE and TM polarizations, within a scanning loss of < 2.7 dB in the measurement. The measured highest SLL of −7.2 dB happens in the 0°-scanning radiation and the maximum gain reaches 20.1 dBi and 21.3 dBi for TE and TM polarization correspondingly. The slight discrepancy between the simulated and measured patterns may be caused by the fabrication tolerances on the unit cells and the beam misalignment between the antenna under test and the reference horn, thus increases the SLL at the boresight and reduce the antenna gain. The maximum gain of the proposed LLA at the boresight over frequency is depicted in Fig. 14(c) for TE and TM polarizations, while the total efficiency reaches 32.3%. It is certain that the antenna loss can be simply reduced and the antenna efficiency can be further improved by replacing the lossy Rogers 4003C substrates with other low-loss substrates in mmW bands. Nevertheless, the proposed design still has a robust dual-polarized scanning performance. TE performs similar to TM due to the high symmetry of the entire structure and the cross-polarization level of less than −24 dB is achieved for the dual polarization. Last but not least, due to the rotational symmetry of the LLA for dual polarization, the scanning performance in *x-o-z* plane which is simply to rotate the flat LL by 90°, is identical to the presented ones in *y-o-z* plane, hence validating its 2D beamscanning capability. Besides, a 2D beam coverage can be alternatively obtained by a 7×7 metasurface planar array extended from the 7×1 linear one.

## 5. CONCLUSION

This paper has theoretically and experimentally validated that with the improved permittivity profile after transformation optics, the impedance mismatch and wave reflection on the flat Luneburg lens surface can be effectively addressed with a free cost on the system weight and size. Combined with a low-profile wideband 7×1 metasurface phased array, a compact dual-polarized millimeter-wave Luneburg lens antenna totally constructed by PCB-based metamaterials and metasurfaces is proposed with a ∼±35° beamscanning coverage and a small *F/D* ratio for two orthogonal polarizations. It achieves a 30.3% impedance bandwidth and the antenna gain is 13.5 dB higher than its feed antenna with the average gain of 7.5 dBi. Therefore, the proposed antenna is a promising solution for the future communications and wireless sensing as an onboard system.